\documentclass{IEEEtran}
\usepackage{cite}
\usepackage{amsmath,amssymb,amsfonts}
\usepackage{graphicx}
\usepackage{textcomp,nicefrac}
\usepackage{textcomp}
\usepackage[switch]{lineno}
\usepackage{xurl}

\def\BibTeX{{\rm B\kern-.05em{\sc i\kern-.025em b}\kern-.08em
T\kern-.1667em\lower.7ex\hbox{E}\kern-.125emX}}
\begin{document}
\title{Operational experience and evolution of the ATLAS Tile Hadronic Calorimeter Read-Out Drivers}
\author{A. Valero, on behalf of the ATLAS Tile Calorimeter Collaboration
\thanks{A. Valero is with Instituto de Física Corpuscular (Universidad de Valencia-CSIC), 46980 Valencia, Spain (e-mail: alberto.valero@cern.ch).}
\thanks{This work was supported by the Ministerio de Ciencia e Innovación under projects RTI2018-094270-B-I00 and RTI2018-100863-B-I00.}
\thanks{Copyright 2020 CERN for the benefit of the ATLAS Collaboration. CC-BY-4.0 license}}
\maketitle
\begin{abstract}
TileCal is the central hadronic calorimeter of the ATLAS experiment at the Large Hadron Collider (LHC).  It is a sampling detector where scintillating tiles are embedded in steel absorber plates. The tiles are grouped forming cells, which are read-out on both sides by photomultiplier tubes (PMTs). The PMT digital samples are transmitted to the Read-Out Drivers (ROD) located in the back-end system for the events accepted by the Level 1 trigger system. The ROD is the core element of the back-end electronics and it represents the interface between the front-end electronics and the ATLAS overall Data AcQuisition (DAQ) system. It is responsible for energy and time reconstruction, trigger and data synchronization, busy handling, data integrity checking and lossless data compression. The TileCal ROD is a standard 9U VME board equipped with DSP based Processing Units mezzanine cards. A total of 32 ROD modules are required to read-out the entire TileCal detector. Each ROD module has to process the data from up to 360 PMTs  in real time in less than 10~\textmu s. The commissioning of the RODs was completed in 2008 before the first LHC collisions. Since then, several hardware and firmware updates have been implemented to accommodate the RODs to the evolving ATLAS Trigger and DAQ conditions adjusted to follow the LHC parameters.
The initial ROD system, the different updates implemented and the operational experience during the LHC Run~1 and Run~2 are presented. 
\end{abstract}

\section{Introduction}
\label{sec:introduction}
\begin{figure}[!h]
\centerline{\includegraphics[width=3.5in]{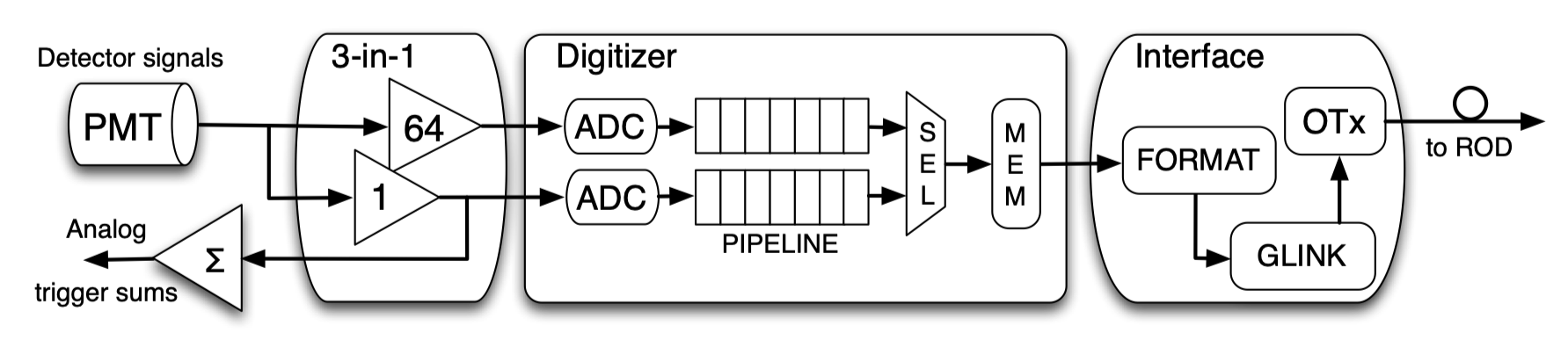}}
\caption{Sketch of the TileCal readout electronics.}
\label{readout}
\end{figure}

TileCal  \cite{Tile} is the central hadronic calorimeter of the ATLAS experiment \cite{ATLAS} at the Large Hadron Collider (LHC).  It is a sampling detector where scintillating tiles are embedded in steel absorber plates. The tiles are grouped forming cells, which are read-out on both sides by photomultiplier tubes (PMTs).  The electric signals produced by the approximately 10000 PMTs are digitized at 40 MHz synchronously with the LHC bunch crossing. The digital samples are stored in pipelined memories located in the front-end electronics \cite{FE}. Seven samples are transmitted to the Read-Out Drivers (ROD)  \cite{ROD} located in the back-end system if the Level 1 (L1) trigger accepts the event (\figurename~\ref{readout}). The commissioning of the RODs was completed in 2008 before the first LHC collisions. Since then, several hardware and firmware updates have been implemented to accommodate the RODs to the evolving ATLAS Trigger and DAQ conditions  \cite{TDAQ} adjusted to follow the evolution of the LHC parameters.

\section{The TileCal read-out driver modules}
\begin{figure*}[!h]
\centerline{\includegraphics[width=6.5in]{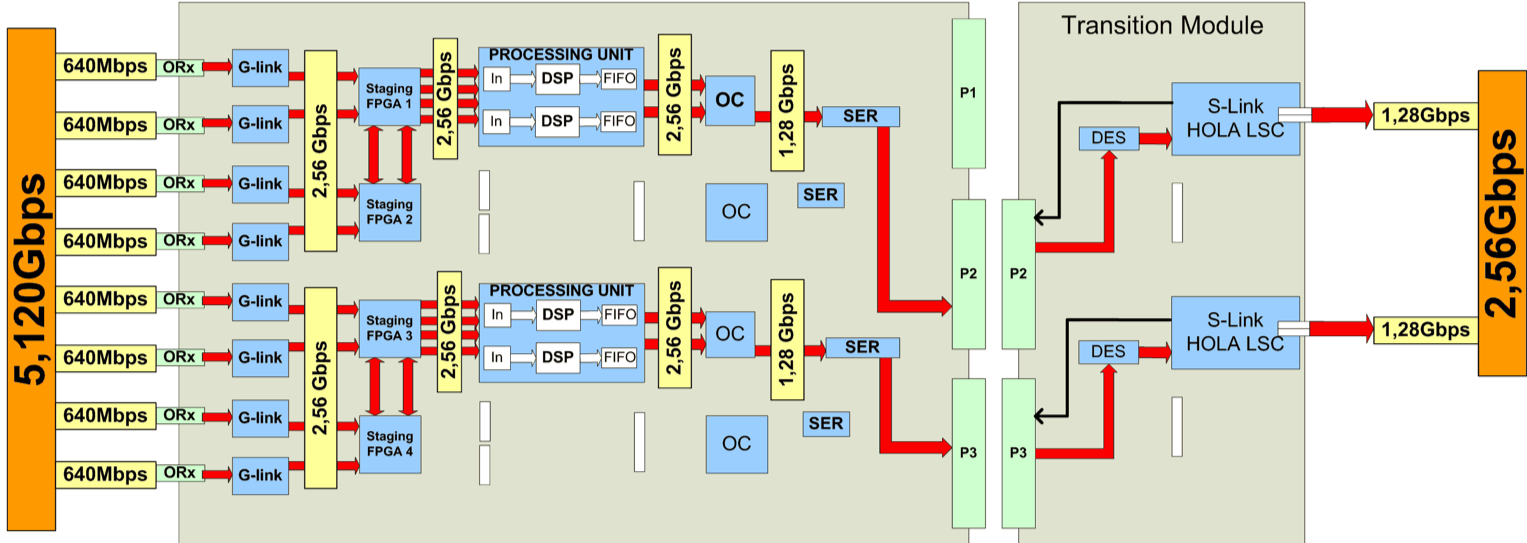}}
\caption{Block diagram of a ROD system equipped with two Processing Units and two HOLA cards.}
\label{ROD}
\end{figure*}

\begin{figure}[!h]
\centerline{\includegraphics[width=3.5in]{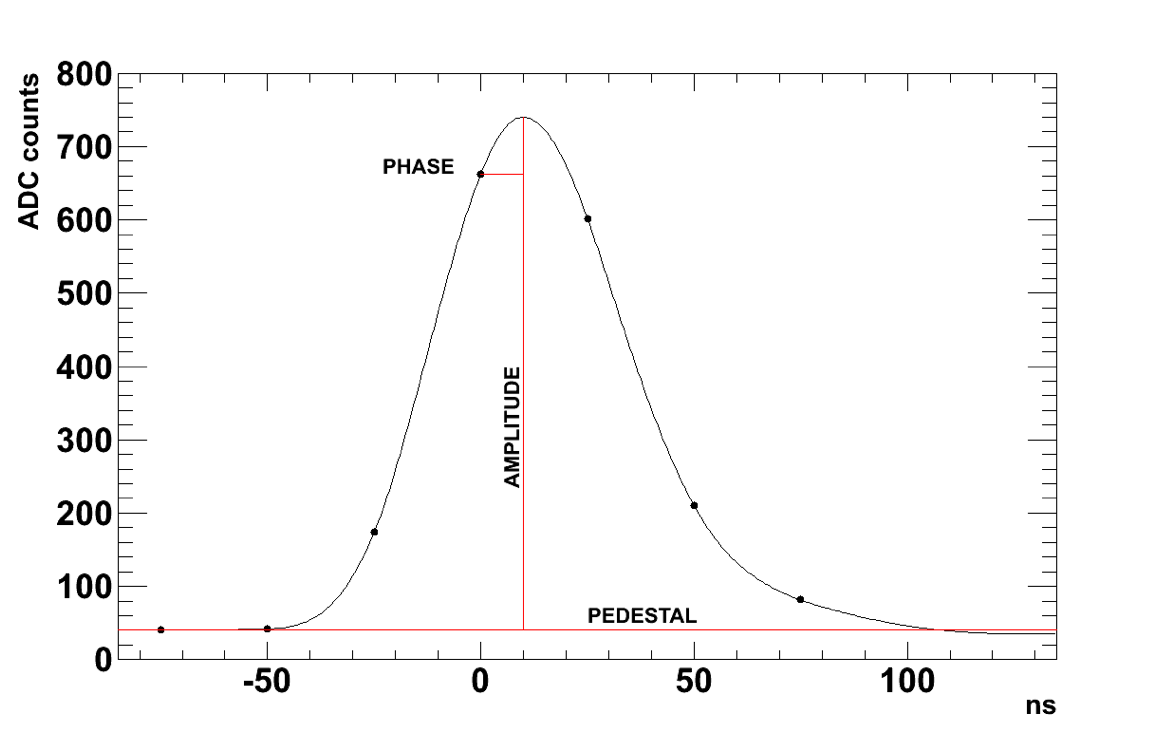}}
\caption{Representation of the TileCal pulse shape with the Optimal Filtering reconstructed magnitudes.}
\label{OF}
\end{figure}

The ROD is the core element of the TileCal back-end electronics. It represents the interface between the front-end electronics and the ATLAS overall Data AcQuisition (DAQ) system. The TileCal ROD module is a standard 9U VME board equipped with DSP based Processing Units mezzanine cards (\figurename~\ref{ROD}) and connected to a Rear Transition Module (RTM) through the VME backplane. 
The RODs are responsible for the energy and time reconstruction, trigger and data synchronization, busy handling, data integrity checking and lossless data compression. 
A total of 32 ROD modules are required to read-out the entire TileCal detector. Each ROD module has to process the data from up to 360 PMTs in real time in less than 10~\textmu s established by the maximum L1 trigger rate \cite{L1}. A digital Finite Impulse Response (FIR) filter called Optimal Filtering (OF) is used in the DSPs to reconstruct the amplitude and phase of the pulses (\figurename~\ref{OF}). It exploits the knowledge of the pulse shape and noise of the electronics and the amount of expected pileup to reduce the contribution of noise and determine the time of deposition.

The input data events and the output-reconstructed fragments are stored in the DSPs in elastic buffers with configurable depth. A veto signal (busy) is generated in the DSP and propagated to the ATLAS Central Trigger Processor to stop the L1 trigger generation when the input buffer is almost full. This mechanism prevents the overwriting of events in the input buffer with the consequent data loss but it introduces undesired deadtime in the detector. The processing time, the output data bandwidth, the depth of the buffers and the almost full flag are the key parameters to reduce the deadtime introduced by the RODs.


\subsection{The Optimal Filtering method}
Optimal Filtering \cite{OF} is a relatively simple algorithm used to reconstruct in real time the energy and phase of the PMT pulses in the DSPs of the RODs.
The OF method relies on the usage of a digitization clock synchronized with the trigger, thus the signal pulses and the samples have always a fixed phase with small variations~\cite{OF}. This feature permits the discrimination of out-of-time pileup energy depositions.

The amplitude and phase of the real pulse is obtained as a linear combination of the digital samples ($S_{i}$) and the weights ($a_{i}$, $b_{i}$).
There are two versions of the OF algorithm; OF1 subtracts the pedestal  ($p$) from the samples before applying the filter (\ref{eq:of1}), whereas OF2 includes an additional constraint in the weights ($\sum_{i=1}^{n}a_{i}=0$) which implies that any common variation in all the samples is cancelled (\ref{eq:of2}):

\begin{equation}
A = \sum_{i=1}^{n}a_{i}(S_{i}-p) \;\;  , \;\;\;  A\tau = \sum_{i=1}^{n}b_{i}(S_{i}-p) ,
\protect\label{eq:of1}
\end{equation}

\begin{equation}
A = \sum_{i=1}^{n}a_{i}S_{i}  \;\;  , \;\;\; A\tau = \sum_{i=1}^{n}b_{i}S_{i} .
\protect\label{eq:of2}
\end{equation}

The weights are computed from the known pulse shape, the expected phase and electronic and pileup noise.
The expected phase and electronics noise is measured with calibration data and they are independent of the LHC conditions.
On the other hand, the pileup noise depends on the LHC beam conditions and the cell position.
Thus, any pulse shape distortion strongly affects the performance of the Optimal Filtering method. In particular, signal pileup different than expected deforms the signal of interest and biases the results of the signal reconstruction.
In addition, there is an iterative version of OF where the weights are selected according to the phase reconstructed in the previous iteration.
This iterative method optimizes the reconstruction when the trigger is not synchronous with the digitizing clock, like in cosmic runs, and when the bunch spacing is larger than the 7 samples window ($\pm$75 ns) which ensures the absence of out-of-time pileup.

\section{Evolution of data processing algorithms during Run~1}

The first LHC beams collided in ATLAS in 2009. Since then, the LHC parameters like bunch spacing and number of interactions per bunch crossing evolved, gradually reaching 75\% of the nominal instantaneous luminosity in 2012. 
The TileCal data acquisition system followed this evolution to optimize its performance and efficiency.
The read-out format and the reconstruction algorithms were optimized with various firmware upgrades.  

\subsection{Commissioning and early Run~1 operation}
During the first year the LHC was operated with larger bunch spacing and lower instantaneous luminosity than the nominal design parameters \cite{TileReady}. Under these conditions the ATLAS L1 trigger rate was reduced, which permitted the usage of large time consuming reconstruction algorithms and  big output data fragments in the RODs with a data taking efficiency close to 100\%.
The iterative version of the OF algorithm was used to minimize the effect of large pulse shifts. This method is capable of detecting the peak of the pulse and apply different filter weights according to its position. 
The output data fragment included both the online reconstructed magnitudes and the front-end digital samples for all the channels. Then, the digital samples were reconstructed offline to validate and certify the online reconstructed magnitudes.

\subsection{ROD firmware optimizations and performance during Run~1}
\begin{figure}[!h]
\centerline{\includegraphics[width=3.5in]{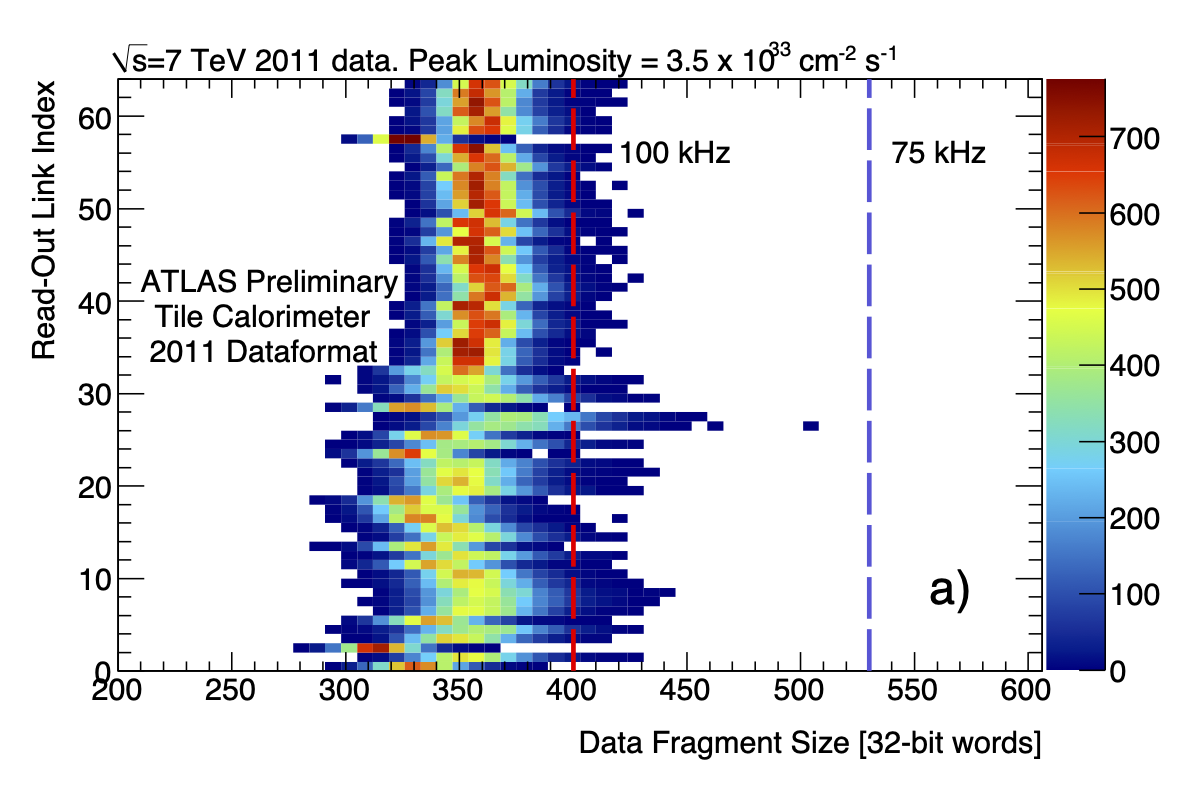}}
\centerline{\includegraphics[width=3.5in]{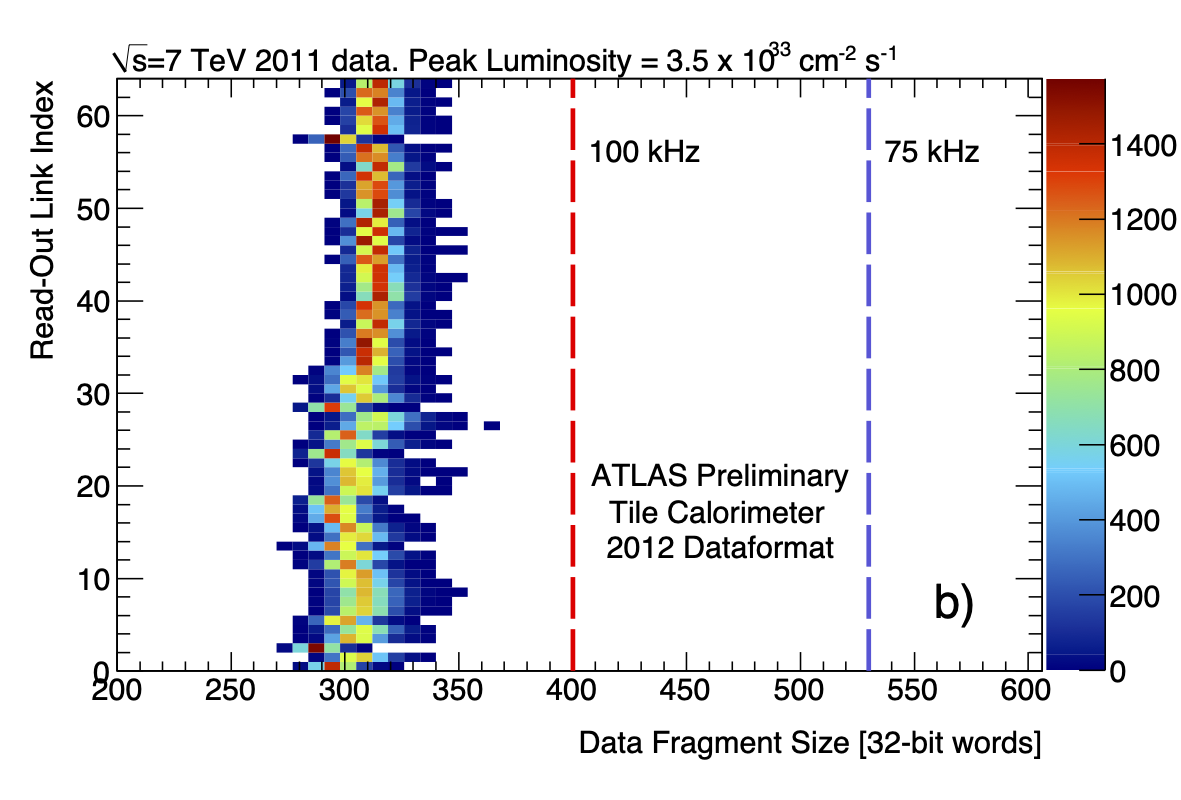}}
\caption{ROD output fragment size per Read-Out Link for the old (a) and new (b) dataformat~\cite{Thesis}.}
\label{dataformat}
\end{figure}

During the second half of Run~1 there was an increase of the LHC instantaneous luminosity and a reduction of the bunch spacing and the consequent increase in the L1 trigger rate forced a change in the read-out strategy. 
In order to minimize the effect of out-of-time pileup and to reduce the computing time, the OF iterative method was replaced by a non-iterative algorithm using pre-calibrated filter weights for each channel.
The increase in the L1 trigger rate reduced the output bandwidth available for each event fragment. A lossless data compression algorithm was implemented to reduce the output fragment size to avoid link saturation and the resulting undesirable deadtime. \figurename~\ref{dataformat} shows the ROD output fragment size for each of the 64 ROD output Read-Out Links (ROL) (32 RODs, 2 ROL each) for a 2012 run with a peak luminosity of $3.5\times 10^{33}$ cm$^{-2}$s$^{-1}$ for the legacy (a) and the compressed (b) output dataformats. The plots show the limit for a free of deadtime operation for a L1 trigger rate of 100~kHz and 75~kHz. The legacy dataformat (\figurename~\ref{dataformat}a) exceeds the limit for 100kHz and thus the data taking efficiency was affected. With the compressed dataformat (\figurename~\ref{dataformat}b) it is possible to run at 100 kHz without deadtime from the RODs.
The data compression algorithm relies on the fact that the majority of the channels have a pedestal-like structure and only few bits are needed to pack the samples.

ATLAS recorded in Run~1 21.3~fb$^{-1}$ of data with a data taking efficiency of 93\%. After these optimizations, TileCal data taking efficiency at the end of Run~1 was around 95\%, mainly due to failures in the front-end low voltage power supply system. The TileCal data quality efficiency was 98.7\%~\cite{TileRun1}.

\section{ROD system upgrade for Run~2}

In view of the expected LHC instantaneous luminosity increase for Run~2, the different ATLAS sub-systems used the Long Shutdown 1 (2013--2014)  to consolidate and upgrade their read-out systems. In particular, TileCal repaired and consolidated the front-end electronics components and replaced the low voltage power supplies with a more radiation tolerant version. As presented in the next sections, the ROD system was also upgraded to operate with the new LHC conditions.
The LHC Run~2 started in 2015 reaching a peak luminosity of two times the nominal value ($2\times 10^{34}$ cm$^{-2}$s$^{-1}$ ) in 2016. The number of proton--proton interactions per bunch crossing exceeded the nominal values thus producing very high detector occupancy, signals pileup and high L1 trigger rates.

\subsection{Motivation for a ROD upgrade}

\begin{figure*}[!h]
\centerline{\includegraphics[width=6.5in]{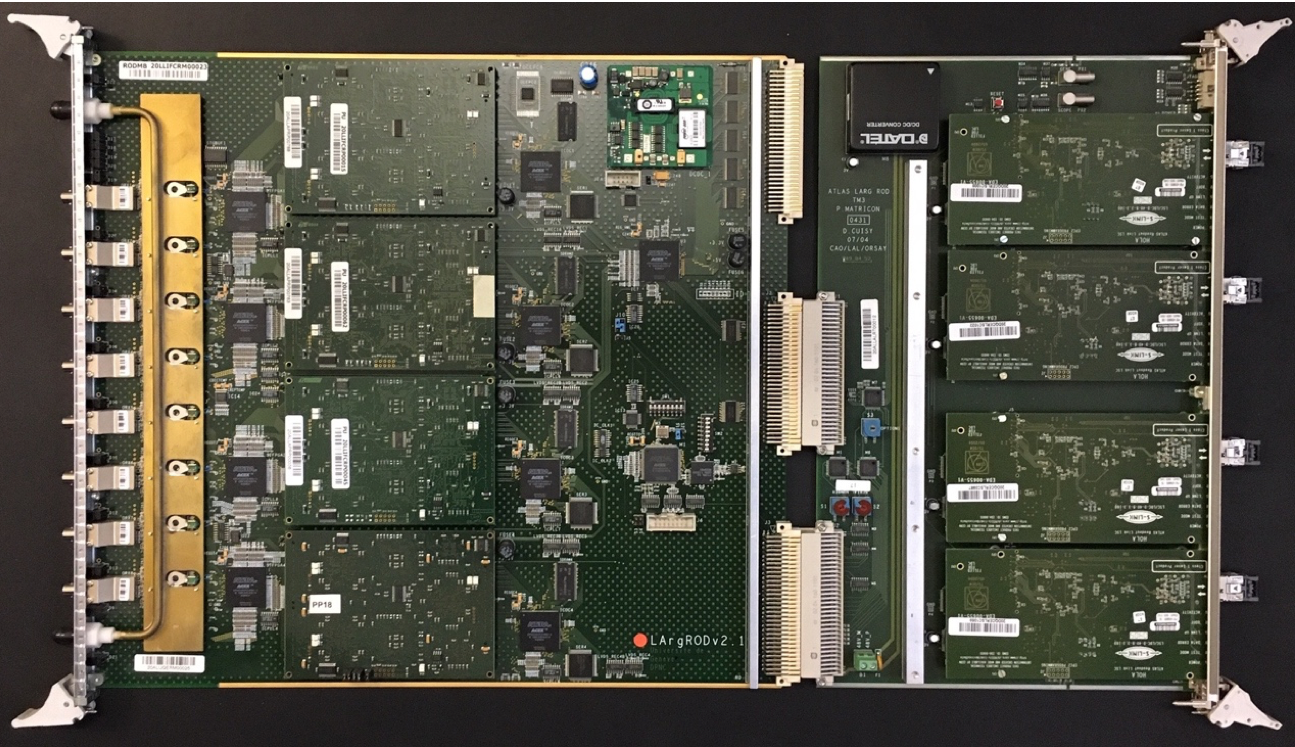}}
\caption{Picture of the upgraded ROD system (ROD motherboard and RTM) equipped with 4 Processing Units and 4 HOLA cards.}
\label{RODupgrade}
\end{figure*}

During the Long Shutdown 1 TileCal undertook a major hardware upgrade in the ROD system in order to cope with the expected evolution of the LHC parameters and in particular the increase of the ATLAS L1 trigger rate while keeping a high data taking efficiency. The ROD processing power and the output data bandwidth were doubled. 
Two Processing Units and two output High-speed Optical Link for Atlas (HOLA) cards per ROD were installed in the available empty slots (\figurename~\ref{RODupgrade}).
A total of 64 HOLA cards dismantled from the ATLAS Muons system were installed in the ROD RTMs. 
The Processing Units cards were produced and certified in the laboratory test-bench during the first year of the Long Shutdown 1. 
The second year of the Long Shutdown 1 was used to install  the 64 Processing Units in the empty ROD slots and make the commissioning of the new upgraded ROD system.
The ROD firmware and data acquisition control software was adapted to run with the new hardware conditions. 

\subsection{ROD upgrade and performance during Run~2}

The signal reconstruction algorithm was updated to avoid a feature of the OF2 method which reconstructs negative energies for large out-of-time pulses.
The increase of the pileup above the nominal value enlarged this effect which was specially affecting high level trigger algorithms based on the  total transverse energy in the calorimeter. Thus, the OF2 method was replaced with OF1 which requires a periodical pedestal calibration but  does not provide negative energies when it is properly measured.
In addition, the pileup noise was introduced in the calculation of the OF weights. Since the pileup noise depends on the cell position, different sets of weights per cell are now used and the database and procedure for loading the values into the DSPs were updated accordingly.

These ROD hardware upgrades and firmware optimizations permitted a successful operation during Run~2 and it should provide stable and smooth operation during the coming Run 3 that should start in 2021 after the Long Shutdown 2 (2019--2020). The ATLAS overall data taking efficiency was 94\% during Run~2 and the data quality losses from TileCal were below 0.4\%.

\section{Conclusions and prospects for Run 3 and beyond}

The TileCal ROD system has evolved following the requirements imposed by the evolution of the LHC during Run~1 and Run~2.
Overall the TileCal data acquisition system has operated successfully during this period with data taking and quality efficiencies compatible with the rest of the ATLAS sub-systems. Moreover, the ROD system is ready to operate smoothly during Run 3 without major modifications.

\begin{figure}[!h]
\centerline{\includegraphics[width=3.5in]{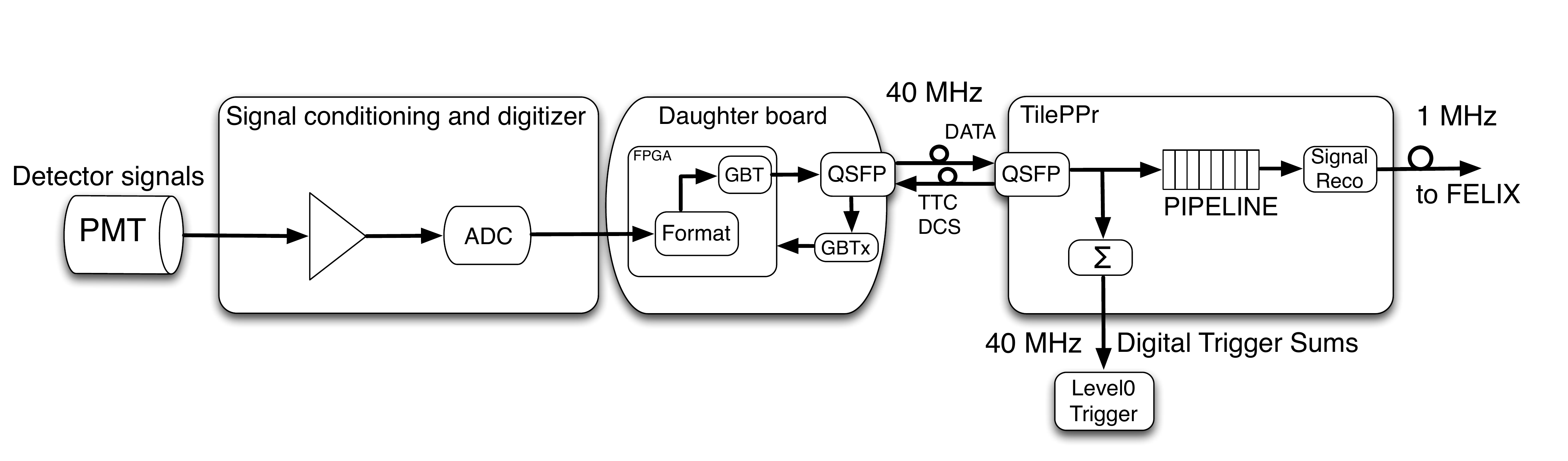}}
\caption{Sketch of the TileCal readout electronics for the HL-LHC.}
\label{HLLHCreadout}
\end{figure}

The ATLAS read-out strategy will be radically changed for the High Luminosity LHC where an increase of up to 7.5 times the LHC nominal instantaneous luminosity is expected. TileCal will replace the read-out electronics and most of the front-end functionalities will be moved to the off-detector PreProcessor (PPr) modules which are the natural evolution of the current RODs \cite{TDRP2} . The PMT signals will be digitized and transmitted to the PPr modules before any event selection is applied (\figurename~\ref{HLLHCreadout}). The PPr will be the interface with the trigger and the ATLAS overall data taking system \cite{TDRP2}.

\end{document}